\documentstyle [12pt]{article}
\makeatletter
\def\eqnarray{%
  \stepcounter{equation}%
  \let\@currentlabel=\theequation
  \global\@eqnswtrue
  \global\@eqcnt\z@
  \tabskip\@centering
  \let\\=\@eqncr
  $$\halign to \displaywidth\bgroup\@eqnsel\hskip\@centering
  $\displaystyle\tabskip\z@{##}$&\global\@eqcnt\@ne
  \hfil$\displaystyle{{}##{}}$\hfil
  &\global\@eqcnt\tw@$\displaystyle\tabskip\z@{##}$\hfil
  \tabskip\@centering&\llap{##}\tabskip\z@\cr}
\makeatother
\title{
\hspace{3.0truein}{\small IFT-488-UNC}\\
\vspace{-0.1truein}
\hspace{3.0truein}{\small NUP-A-94-3}\\
\vspace{0.2truein}
{Spin Transmutation in $(2+1)$ Dimensions}}
\author{
Wei Chen\footnotemark[1]\\
   Department of Physics\\
University of North Carolina\\
Chapel Hill, NC 27599-3255\\
\\
Chigak Itoi\footnotemark[2]\\
Department of Physics and Atomic Energy Research Institute\\
College of Science and Technology, Nihon University\\
Kanda Surugadai, Chiyoda-ku, Tokyo 101, Japan
}
\date{
}
\footnotetext{$^*$ chen@physics.unc.edu}
\footnotetext{$^\dagger$  itoi@phys.cst.nihon-u.ac.jp}

\begin{document}
\maketitle
\vspace{0.2truein}
\begin{abstract}
We study a relativistic anyon model with
a spin-$j$ matter field minimally coupled to
a statistical gauge potential governed by the Chern-Simons
dynamics with a statistical parameter $\alpha$.
A spin and statistics transmutation is shown in terms of
a continuous random walk method.
An integer or odd-half-integer part of
$\alpha$ can be reabsorbed by change of $j$.
 We discuss the equivalence of a large class of
(infinite number) Chern-Simons matter models
for given $j$ and $\alpha$.

\end{abstract}
\newpage
\baselineskip=22.0truept

\renewcommand{\theequation}{\thesection.\arabic{equation}}
\baselineskip=18.0truept
\newpage
\vskip 0.5truein
\section{Introduction}
\setcounter{equation}{0}
\vspace{3 pt}

It is well-known now that in (2+1) spacetime dimensions there are
anyons which have arbitrary spin and statistics
\cite{M}.The statistics of particles is
changed if an interaction with a statistical
gauge potential governed by the
Chern-Simons term is introduced
\cite{H}. While this issue is settled for non-relativistic
matters, there has been debate as far as the correct
formalism of relativistic anyons is concerned.
In this context, the spin degree of freedom plays an
essential role.
In the framework of quantum field theory, a
Fermi-Bose transmutation has been observed
in a (2+1) dimensional system \cite{P}. Namely that
a charged scalar field coupled to a Chern-Simons
gauge potential at a specific value of
Chern-Simons coefficient (or statistical parameter)
turns out to be a free fermion field theory.
In this example, one has to consider
spin transmutation
as well as statistics transmutation.
These refer to apparently different objects:
a statistical transformation affects
the Aharonov-Bohm phase of the wave-function
of two identical particles
moving around one to the other on a plane,
while any spin transmutation should cause
corresponding change in the
Lorentz group representation of the matter fields.
Though there was investigation on anomalous
magnetic momentum
in a perturbation
theory in the literature \cite{KS},
a clear understanding for
spin transmutation seems still lacking.

The problem with spin transmutation
is discussed in a recent letter \cite{CI}.
The present paper is to continue the discussion.
We see that in a general Chern-Simons
matter field theory model, any integer or
odd-half-integer part of the
Chern-Simons coefficient can be reabsorbed
by changing the character of
the Poincar$\acute{{\rm e}}$
representation describing the matter.
This is an observation through a continuous random walk
method with certain assumptions with respect to a
regularization. In this formalism,
the partition function of an anyon system
can be represented in particle path integrals with some
phase factor with terms including a spin factor from
the path integral of the matter field and
a self-energy and a relative energy from that
of the Chern-Simons gauge field. And,
a topological relation between the spin factor
and self-energy enables us to show
the spin and statistics transmutation.
Moreover, we see that, due to the quantization
of the Chern-Simons
relative energy, only integer or odd-half-integer part of
the coupling constant endows the anyons with
a change of spin and a ``small'' part of it
($<1/2$) remains representing a residual
Chern-Simons interaction.

There is a nice application with the spin and
statistics transmutation. As we know,
the Chern-Simons coefficient is used quite often
as a parameter over which a perturbation expension
is conducted. However, when the Chern-Simons
coefficient is required large in some systems,
the perturbation expension is not reliable. Now,
with the spin and statistics transmutation,
one can trade a large part of the Chern-Simons coefficient
for a change of the spin of the matter and keep the
remaining Chern-Simons coupling weak.
In this way, a perturbation
expansion becomes controllable.

As in any relativistic field theory, the fundamental
fields are required to carry irreducible unitary
representations of the universal covering group of
Poincar$\acute{{\rm e}}$ group.
One way to realize this is to construct the fundamental
fields transforming as linear representations of the
Lorentz group, and to subject them to constraints
that eliminate unphysical degrees of freedom.
A striking feature of the Poincar$\acute{{\rm e}}$ representations
in ($2+1$) dimensions is that each of them has only one
independent component. This can be understood with
the following argument. Define the value of spin $j$,
which can be arbitrary in ($2+1$) dimensions \cite{Sch}, via
 $j^\mu j_\mu = j(j+1)$ with
$j^\mu$ the spin part of the Lorentz generators.
To construct an irreducible linear representation,
one needs $2j+1$ components for an integer or
odd-half-integer spin-$j$. However, a constraint,
the Pauli-Lubanski condition, kills all but one
component.

Like statistics transmutation, spin transmutation
is obviously a phenomenon at the quantum level.
A key step to understand the spin transmutation
in a Chern-Simons matter model is
to establish a relation between the spin $j$ and
the Chern-Simons coefficient
$\alpha$, while a
place we feel convenient to do so is
the partition function of an
anyon system, where the matter fields and gauge
potential are all integrated out.
 From proposing an anyon model with an arbitrary spin-$j$
spinning matter field minimally coupled to a Chern-Simons
gauge potential (a local four-body matter
self-interaction is also introduced),
we offer a step-by-step analysis
and obtain a $j$-$\alpha$
relation that displays the spin transmutation.
We then consider some consequences with emphasis on
the equivalence of a large
class of (infinite number) Chern-Simons matter models.
We also shed lights on the debate
about the role a Chern-Simons gauge field plays,
and on the possible role the Chern-Simons spinning
field theory model plays in describing the critical
phenomena.

In section 2, we briefly review the
Poincar$\acute{{\rm e}}$
group in $(2+1)$ dimensions and its representations,
with details for $j=1/2, 1$, and $3/2$. In section 3,
we define a spin factor
for matter fields, and introduce $SU(2)$ coherent
states to formulate the
free-energy of spin-$j$ matter fields,
treating the Chern-Simons
gauge potential as an external field.
We then carry out, in section 4, the functional
integral over the Chern-Simons field;
we analyze the relative and self-energies
from the Chern-Simons gauge potential and verify
their relations to a topological quantity,
the Gaussian linking number,
and to the spin factor associated with the
spin of matter fields.
 From these the spin transmutation is explicitly observed.
We discuss the equivalence of a sort of Chern-Simons
matter models with a local four-body
interaction in section 5,
while concluding remarks are given in section 6.

\vskip 0.5truein
\section{Poincar$\acute{{\rm\bf e}}$ Group and Representations}
\setcounter{equation}{0}
\vspace{3 pt}

The three-dimensional Poincar$\acute{{\rm e}}$ group $\pi$
is defined as the set of real transformations
\begin{equation}
(a, \Lambda):~~x^\mu  \rightarrow
 \Lambda_{\nu}^{~\mu}x^\nu + a^\mu
\end{equation}
in the three-dimensional Minkowski space which leave
$g_{\mu\nu}(x-y)^\mu (x-y)^\nu$ 
invariant, where the metric $g_{\mu\nu} = diag(1,-1,-1)$.
The group $\pi$ is actually a semidirect product of the
translation and Lorentz groups, $N$ and $L$,
as applying two successive transformations on $x$, one has
$(a^\prime, \Lambda^\prime)(a, \Lambda)
= (a^\prime + \Lambda^\prime a, \Lambda^\prime\Lambda)$.

The Hermitean generators
of (infinitesimal) Lorentz transformations
$L_{\mu\nu}$, realized as
$i(x_\mu\partial_\nu-x_\nu\partial_\mu)$
with $\partial_\mu = \partial/\partial x^\mu
= (\partial_0, {\bf \nabla})$, obey
the Lie algebra of $SO(2,1)$
\begin{equation}
[L_{\mu\nu},L_{\rho\tau}] =
i(g_{\mu\tau}L_{\nu\rho}-g_{\nu\tau}L_{\mu\rho}
+g_{\mu\rho}L_{\nu\tau}-g_{\nu\rho}L_{\mu\tau})\;.
\label{alg1}
\end{equation}
The most general representation of the generators
of $SO(2,1)$
satisfying (\ref{alg1}) is
\begin{equation}
 M_{\mu\nu} = i(x_\mu\partial_\nu-x_\nu\partial_\mu)
+ S_{\mu\nu}\;,
\end{equation}
where the Hermitean operators $S_{\mu\nu}$, introduced
for describing spin of particles and thus called spin matrix,
satisfy the same Lie algebra as $L_{\mu\nu}$ and commute with them.
Formally, the discussion so far looks similar to
that in the $(d+1)$ dimensions for $d > 2$.
But there is an essential difference, $i.e.$
in (2+1) dimensions, the little group
$M_{ij}~ (i,j = 1,2)$ of the Lorentz rotation
is Abelian ($SO(2)$) and moreover the universal covering
group of $SO(2)$ can be imbedded in the universal
covering group of $SO(2,1)$, and thus
its eigenvalue can be any real number \cite{Pl}.
Therefore spin of particles in $(2+1)$ dimensions
can be arbitrary, while
spin of particles in higher dimensions must be quantized.

Introducing a new operator
\begin{equation}
J^\mu= \frac{1}{2}\epsilon^{\mu\sigma\lambda}
M_{\sigma\lambda}\;,
\label{J}
\end{equation}
the Lorentz algebra takes a form
\begin{equation}
[J^\mu , J^\nu]  =
i \epsilon^{\mu\nu\lambda}J_\lambda \;.
\label{alg2}
\end{equation}

The Hermitean generators of translations
$P_\mu$, realized as  $i \partial_\mu$, satisfy
\begin{eqnarray}
\left[ P_{\mu}, P_{\nu} \right] &=& 0\; , \label{alg3} \\
\left[ J^{\mu}, P^{\nu} \right]
&=& i \epsilon^{\mu \nu \lambda}P_\lambda \;. \label{alg4}
\end{eqnarray}
(\ref{alg2})-(\ref{alg4}) form
the three-dimensional Poincar$\acute{{\rm e}}$ algebra.

As in any field theory,
the matter fields that describe the elementary
particles are normally required to
carry irreducible unitary representations (UIR's) of
Poincar$\acute{{\rm e}}$ group, so that the
particles are indivisible and the probability
amplitudes calculated from the theories
are invariant. A complete set of
UIR's of the three-dimensional
Poincar$\acute{{\rm e}}$ group is given in \cite{B}.
Alternatively, one may begin with covariant fields,
fields transforming as
\begin{equation}
(a,\Lambda):~~\Phi(p) \rightarrow e^{ip\cdot a}
D(\Lambda)\Phi(\Lambda^{-1}p)\;,
\label{trans}
\end{equation}
in the momentum space where
$P^\mu$ is replaced by the eigenvalue $p^\mu$.
The exponential $e^{ip\cdot a}$ in (\ref{trans})
is the UIR of the translation group $N$,
and $D(\Lambda)$ is an appropriate
representation of the Lorentz group.
Since a covariant
field is in general not irreducible,
one subjects the covariant field to constraint(s)
to remove unphysical degrees of freedom.
This latter approach that
we shall follow is convenient for constructing
interactions in local terms.

In the (2+1) dimensional
Poincar$\acute{{\rm e}}$ algebra, the invariants,
or the  Casimir operators
that commute with all the generators, are
$P^2$ and the Pauli-Lubanski scalar $P\cdot J$.
The spin part of $J_\mu$, denoted as
$j_\mu$, defines the value of spin $j$ via
$j^\mu j_\mu = j(j+1).$
$P^2$ gives the mass-shell condition,
\begin{equation}
[P^2 - (j/k)^2M^2]\Phi = 0\;,
\label{cons1}
\end{equation}
where k ($0< k\leq j$) labels the non-zero eigenvalues
of $j_\mu$, for any given spin-$j$.
Apparently the mass spectrum now
is $\pm jM/k$ for all $k$ \cite{note6}.
And the Pauli-Lubanski scalar provides an
additional constraint 
\begin{equation}
(P\cdot J +jM)\Phi = 0\;, \label{cons2}
\end{equation}
which specifies
the helicity,
as the Pauli-Lubanski
scalar $P\cdot J$ concerns no
orbit angular momentum
($P_\mu\epsilon^{\mu\nu\lambda}
L_{\nu\lambda} = 0$).

By these constraints, one can construct
linear representations
of the Poincar$\acute{{\rm e}}$ group.
For an integer or odd-half-integer $j$,
$2j+1$ components are usually required for
the wave-function $\Phi$.
However, as the Pauli-Lubanski condition
consists of $2j+1$ homogeneous equations
for the $2j+1$ components, and as a
$(2j+1)\times(2j+1)$ Pauli-Lubanski matrix
$P\cdot J+jM$ is of rank $2j$ under a mass-shell
condition,  one and only one of these components
is independent.
For an arbitrary $j$, infinite number components
are needed for a linear representations,
while constraints kill all but one component \cite{JN}.
Moreover, for a given $j$ ($>1$ if $j$ is an integer
or odd-half-integer), there are more than
one such wave-function characterized by $k$.
All these span a complete basis of the spin-$j$
irreducible representation.

For the later use, in the rest of this section,
we consider $j = 1/2, 1,$ and $3/2$ as examples.
For spin-$1/2$ particles,
it is convenient to choose
\begin{equation}
j^\mu = \frac{1}{2}\gamma^\mu\;,
\label{j1}
\end{equation}
where $2\times 2$ matrices
$\gamma^\mu = (\sigma^3, -i\sigma^1, i\sigma^2)$,
with $\sigma^1, \sigma^2$ and $\sigma^3$ the
Pauli matrices.
$j^\mu$ obey the Lorentz algebra (\ref{alg2}).
The two-component Dirac field $\psi$ transforms as
\begin{equation}
\Lambda:~~\psi(p) \rightarrow
e^{i\omega_\Lambda\cdot j}\psi(\Lambda^{-1}p)\;.
\end{equation}
The Pauli-Lubanski
condition (\ref{cons2}) then is precisely the Dirac
equation:
\begin{equation}
i\gamma \cdot \partial \psi + M\psi = 0\;.
\label{eq1}
\end{equation}
Above the Dirac indefinite scalar product
is used so that (\ref{j1}) is self-conjugated and
the representation of the Poincar$\acute{{\rm e}}$
group is unitary.
The (positive energy) solution of (\ref{eq1}) in
the momentum space  is
\begin{equation}
\psi(p) = \frac{1}{\sqrt{2M(E+M)}}
\left(\begin{array}{c}
p^y-ip^x\\E+M\end{array}\right)\phi(p)\;,
\label{psi}
\end{equation}
where $E = \sqrt{{\mbox{\boldmath $p$}}^2 + M^2}$ and
the normalization is so chosen that,
in the rest-frame $\acute{p}^\mu = (M,0,0)$,
$\psi^\dagger\psi = \phi^*\phi$, with $\phi(p)$
a scalar function.
(\ref{psi}) provides the spin-$1/2$ representation.
Obviously, there exist \cite{JN}
a Lorentzian transformation that boosts $p^\mu$ from its rest-frame
$\acute{p}^\mu$ and an associated transformation
that boosts the solution (\ref{psi}) from
$\left(\begin{array}{c}
0\\1\end{array}\right)\phi(\acute{p})$.

For spin-1 particles, the covariant vector field $B_\mu$ transforms
as
\begin{equation}
\Lambda:~~B_\mu(p) \rightarrow \Lambda_\mu^{~\nu}B_\nu(\Lambda^{-1}p)\;.
\end{equation}
Accordingly, the generators $j^\mu$ can be chosen as
\begin{equation}
(j^\mu)_{\sigma\lambda} =
-i\epsilon^\mu_{~\sigma\lambda}\;,
\label{j2}
\end{equation}
which satisfy the Lorentz algebra (\ref{alg2}). Acting on the three-vector
$B^\mu$, the Pauli-Lubanski condition
(\ref{cons2}) with $j=1$ is
\begin{equation}
-\epsilon^{\mu\nu\lambda}\partial_\nu B_\lambda +M B^\mu
= 0\;.
\label{eq2}
\end{equation}
It is not difficult to check that
the on-shell solution of (\ref{eq2})
is
\begin{equation}
B^\mu(p) = \frac{1}{\sqrt2}
\left[\left(\begin{array}{c}
0\\1\\i\end{array}\right)
+\frac{p^x+ip^y}{M(E+M)}\left(\begin{array}{c}
E+M\\p^x\\p^y\end{array}\right)\right]
\phi(p)\;.
\label{B}
\end{equation}

(\ref{eq2}) is actually
the equation of motion of a field theory described by
the Lagrangian
\begin{equation}
\frac{1}{2}B_{\mu}
[-\epsilon^{\mu\nu\lambda}\partial_\nu
+Mg^{\mu\lambda}]B_\lambda\;.
\label{SB0}
\end{equation}

On the other hand, in the present work
we treat the vector $B_\mu$ as a fundamental charged
spin-$1$ matter field, and couple it minimally
to a Chern-Simons gauge potential $a_\mu$ and an
external gauge fields $C_\mu$. Namely, we consider
the local gauge invariant Lagrangian
\begin{equation}
{\cal L}_1 =
\frac{1}{2}B^*_{\mu}
[-\epsilon^{\mu\nu\lambda}(\partial_\nu+ia_\nu+iC_\nu)
+Mg^{\mu\lambda}]B_\lambda + {\rm ~CS~term~of}~a_\mu \;.
\label{SB1}
\end{equation}
The corresponding $U(1)$ gauge transformations are
\begin{equation}
a_\mu \rightarrow a_\mu -  \partial_\mu \epsilon_1\;,~~~
C_\mu \rightarrow C_\mu -  \partial_\mu \epsilon_2\;,~~~
B_\mu  \rightarrow e^{i(\epsilon_1+\epsilon_2)}B_\mu\;.
\end{equation}
The dynamics of the gauge field
$a_\mu$ is governed by the Chern-Simons term.
Due to the topological nature of Chern-Simons action,
$a_\mu$ field carries no independent degree of freedom.
As we shall explicitly see in the following sections,
the Chern-Simons interaction changes the spin of matter
field \cite{note1}.

For $j=3/2$, we construct the spin matrix
satisfying
the Lorentz algebra (\ref{alg2})
as the $4\times 4$ matrices
\begin{eqnarray}
& &\frac{1}{2}
\left(\begin{array}{cccc}
3& & & \\
 &1& & \\
& &-1& \\
& & &-3
\end{array}\right)\;,~~~
-\frac{i}{2}
\left(\begin{array}{cccc}
       &\sqrt3 &   & \\
\sqrt3&       &2  & \\
       &     2 &   & \sqrt3\\
       &       &\sqrt3&
\end{array}\right)\;,~~~~
\frac{1}{2}
\left(\begin{array}{cccc}
       &\sqrt3&   & \\
-\sqrt3&       &2& \\
       &-2 &   &\sqrt3\\
       &       &-\sqrt3  &
\end{array}\right)\;,\nonumber\\
& &
\label{j3}
\end{eqnarray}
for $j_0, j_1$ and $j_2$, respectively.
It is readily to check that
$j^2 =(3/2)(3/2+1)$.
The four-component spin-$3/2$ field $\Psi$ is subject to the
mass-shell
  and Pauli-Lubanski conditions. The latter condition
is actually the equation of motion for the field $\Psi$:
\begin{equation}
ij\cdot\partial\Psi + \frac{3}{2} M\Psi = 0\;.
\label{eq3}
\end{equation}
The on-shell positive energy solutions of (\ref{eq3}) in
the momentum space take a form
\begin{equation}
\Psi_k(p) =
\frac{\sqrt3}{\sqrt{2M(E_k+jM/k)}}\left(\begin{array}{c}
(E_k-jM/k)/\sqrt{3}\\
p^y+ip^x\\
p^y-ip^x\\
(E_k+jM/k)/\sqrt{3}\end{array}\right)\phi(p)\;,
\label{Psi}
\end{equation}
where $E_k =\sqrt{{\mbox{\boldmath $p$}}^2 + (jM/k)^2}$,
$j=3/2$, $k=1/2$, or $3/2$. These two states span
a complete basis of spin-$3/2$ irreducible representation.

In the above, we have assumed the mass
parameter $M$ positive and worked out the representations
for spin $j=1/2, 1,$ and $3/2$, respectively.
One may set $M$ as $-|M|$ and keep all the others same,
then the representations constructed will be
for $j=-1/2, -1,$ and
$-3/2$. In other words, with the sign of $j$ fixed,
the sign of mass differs
the spin ``up'' and ``down'' \cite{note3},
corresponding to the two helicity directions.
For concreteness, we assume $j$
positive thereafter.
\vskip 0.5truein
\section{
Path Integral of
Spinning Particles
}
\setcounter{equation}{0}
\vspace{3 pt}

We turn to consider how the Chern-Simons interaction
affects the spin of matter fields.
The sort of Chern-Simons matter models of interest
can be uniformly described by the Lagrangian
\begin{equation}
{\cal L}(j,\alpha) =  \bar\Phi
[j^{-1} j_\mu(\partial_\mu + ia_\mu
+iC_\mu) +M]\Phi
- \frac{i}{8\pi \alpha}
\epsilon_{\mu\nu\lambda}a_\mu\partial_\nu a_\lambda\;.
\label{SJ}
\end{equation}
Besides the mass parameter $M$,
the system is characterized
by two real parameters,
$j$, which we call the spin of the field $\Phi$,
the character of irreducible representations of the
Poincar$\acute{{\rm e}}$ group and $\alpha$
the so-called statistical parameter.
The spin $j$ is defined by $j_\mu j_\mu = j(j+1)$ with
the spin matrix $j_\mu$ obeying the Lorentz algebra
(\ref{alg2}).
The statistical parameter $\alpha$ represents
a change of the statistics of the matter field
brought by the
Chern-Simons interaction. It represents also a change of
the spin of the matter field, as we come to see now.
Viewing (\ref{SJ}) as an interacting field
theory, on the other hand, $\alpha$
reflects the strength of the interaction among
the matter fields and Chern-Simons gauge potential.
To manifest this, one rescales
the Chern-Simons field $a_\mu \rightarrow
\sqrt{4\pi\alpha}a_\mu$ so that the Chern-Simons term
is normalized to $1/2$, and then the gauge coupling
constant is $\sqrt{4\pi\alpha}$.
To be more general, we have introduced an external gauge
field $C_\mu$ also minimally coupled to the matter fields.

Let's calculate the partition function of the Chern-Simons
matter model (\ref{SJ}).
For this purpose, we make a Wick rotation and
come to the Euclidean spacetime.
We have the partition function
\begin{equation}
Z{(j,\alpha)}=\int{\cal D}a_\mu
           {\cal D}\bar{\Phi}
           {\cal D}\Phi
e^{-\int d^3x {\cal L}(j,\alpha)} \;,
\end{equation}
keeping in mind that the partition function is a functional
of the external gauge field $C_\mu$, and
pending a gauge fixing concerning
the gauge freedom with the Chern-Simons field until
the integral over $a_\mu$ field is to be carried out.
With the minimal coupling, the path integrals over
the matter fields and over the Chern-Simons gauge
potential both are of Gaussian type.
We choose to perform the former first.
By a standard method, we obtain
\begin{equation}
Z{(j,\alpha)}=\int{\cal D}a_\mu\sum^\infty_{n=0}
\frac{1}{n!}(-W_j)^n
 \exp[\frac{i}{8\pi\alpha}\int d^3x
\epsilon^{\mu\nu\lambda}
a_\mu\partial_\nu a_\lambda]\;,\label{Z1}
\end{equation}
where  the free energy of the spin-$j$ matter field
$W_j$ is
\begin{equation}
W_j={\rm Tr} \log[-i(\partial_\mu+ia_\mu+iC_\mu)
j_\mu j^{-1} +M]\;. \label{W1}
\end{equation}

To render the remaining integral over
$a_\mu$ in (\ref{Z1})
an explicit Gaussian, we invoke the
path integral representation of spinning
particles \cite{SSS} to formulate $W_j$.
To start, we introduce
the SU(2) spin coherent states in the spin-$j$
representation \cite{AP}, which are parametrized by
a unit vector ${\mbox{\boldmath $e$}}$ in three dimensions
\begin{equation}
|{\mbox{\boldmath $e$}}> \equiv \exp[-i \theta
({\mbox{\boldmath $e$}}_3 \times {\mbox{\boldmath $e$}})
 \cdot
{\mbox{\boldmath $j$}} |{\mbox{\boldmath $e$}}_3
\times {\mbox{\boldmath $e$}} |^{-1}]|0>\;,
\label{CH}
\end{equation}
where
${\mbox{\boldmath $e$}}_3 =(0, 0, 1)$,
 $\theta$ is the angle between ${\bf e}_3$ and ${\bf e}$,
and $|0>$ is the highest weight state
in the spin-$j$ representation. Note, this definition
involves the spin matrix $j_\mu$
with the magnitude $j$.
The spin coherent state has the following three
properties:\\
(1) Unity partition
\begin{equation}
\int d{\mbox{\boldmath $e$}} |{\mbox{\boldmath $e$}}>
<{\mbox{\boldmath $e$}}| =1\;,
\label{PU}
\end{equation}
where $d{\mbox{\boldmath $e$}}$ is the
rotationally invariant measure on
the two dimensional unit sphere. \\
(2) Area law
\begin{equation}
<{\mbox{\boldmath $e$}}+ \delta {\mbox{\boldmath $e$}}
|{\mbox{\boldmath $e$}}>
= \exp \{ ijA[{\mbox{\boldmath $e$}}, {\mbox{\boldmath $e$}}+
\delta{\mbox{\boldmath $e$}}, {\mbox{\boldmath $e$}}_3 ]\}
+ O( \delta{\mbox{\boldmath $e$}}^2)\;,
\label{IP}
\end{equation}
where $A[{\mbox{\boldmath $e$}}, {\mbox{\boldmath $e$}}+
\delta{\mbox{\boldmath $e$}}, {\mbox{\boldmath $e$}}_3 ]$
is the area of
a spherical triangle with the vertices
${\mbox{\boldmath $e$}}$, ${\mbox{\boldmath $e$}}+
\delta{\mbox{\boldmath $e$}}$ and ${\mbox{\boldmath
$e$}}_3$. \\
(3) Expectation value
\begin{equation}
 <{\mbox{\boldmath $e$}}|~{\mbox{\boldmath
$j$}}~|{\mbox{\boldmath $e$}}>=j {\mbox{\boldmath $e$}}\;.
\label{EV}
\end{equation}
With these properties, 
the trace Tr can be represented
in terms of path integral over
random paths  that are closed
on a two dimensional sphere, such as
\begin{eqnarray}
{\rm Tr}[ Texp\left( \int_0^L dt
{\mbox{\boldmath $j$}}\cdot{\mbox{\boldmath $S$}}(t)\right)]
&\equiv& \lim_{N \to \infty}
{\rm Tr} \prod_{i=1}^N \left( 1+
\Delta t {\mbox{\boldmath $j$}}\cdot
{\mbox{\boldmath $S$}}(t_i)\right) \nonumber \\
&=& \int {\cal D}{\mbox{\boldmath $e$}}
\exp \left( ij \Xi [{\mbox{\boldmath $e$}}]
+ j \int_0^L d{\mbox{\boldmath $e$}} \cdot
{\mbox{\boldmath $S$}} \right)\;,
\label{TR}
\end{eqnarray}
where ${\mbox{\boldmath $S$}}(t)$ is a c-number source, and
\begin{equation}
 \Xi[{\mbox{\boldmath $e$}}] = \int_D dt ds{\mbox{\boldmath
$e$}}\cdot
[\partial_s{\mbox{\boldmath $e$}}\times
\partial_t{\mbox{\boldmath $e$}}]
\label{phi}
\end{equation}
is defined as spin factor.
${\mbox{\boldmath $e$}}(t,s)$ is an extention
of ${\mbox{\boldmath $e$}}(t)$ satisfying
${\mbox{\boldmath $e$}}(t,0)=$ const and
${\mbox{\boldmath $e$}}(t,1)={\mbox{\boldmath $e$}}(t)$.
Geometrically, the spin factor
is the area enclosed by the closed path ${\mbox{\boldmath
$e$}}(t)$
($0\leq t \leq L$)
on the two dimensional unit sphere.

Introducing Schwinger parameter $L$,
we write the free energy of spin-$j$ particles
$W_j$ as
\begin{eqnarray}
W_j&=&{\rm Tr} \log[(\partial_\mu
+ia_\mu+iC_\mu)j_\mu
 j^{-1} +M] \nonumber \\
&=& \int_{\epsilon}^{\infty}
\frac{dL}{L} {\rm Tr}
e^{-L\left( i({\mbox{\boldmath $p$}}
+ {\mbox{\boldmath $a$}} + {\mbox{\boldmath $C$}}) \cdot
{\mbox{\boldmath $j$}}j^{-1}
+ M \right)}\;,
\end{eqnarray}
where $\epsilon$ is an ultraviolet cutoff. Using the $SU(2)$
coherent states, we have
\begin{equation}
W_j=\int_{\epsilon}^{\infty} \frac{dL}{L}
\int  \prod_{i=1}^{N}
d{\mbox{\boldmath $x$}}_i d{\mbox{\boldmath $e$}}_i
<{\mbox{\boldmath $x$}}_i|
<{\mbox{\boldmath $e$}}_i|
e^{-\Delta L\left( i({\mbox{\boldmath $p$}}
+ {\mbox{\boldmath $a$}}
+ {\mbox{\boldmath $C$}})
 \cdot {\mbox{\boldmath $j$}}j^{-1}
+ M \right)}|{\mbox{\boldmath $e$}}_{i-1}>|{\mbox{\boldmath
$x$}}_{i-1}>.
\label{SL}
\end{equation}
The path integrals over ${\mbox{\boldmath $x$}}_i$
and ${\mbox{\boldmath $e$}}_i$
are subject to periodic boundary
conditions ${\mbox{\boldmath $x$}}_N={\mbox{\boldmath $x$}}_0$
and ${\mbox{\boldmath $e$}}_N ={\mbox{\boldmath $e$}}_0$.
The infinitesimal kernel in terms of the coherent states
and the momentum eigenstates is
\begin{eqnarray}
 & &<{\mbox{\boldmath $x$}}_i|<{\mbox{\boldmath $e$}}_i|
e^{-\Delta L\left( i({\mbox{\boldmath $p$}}
+ {\mbox{\boldmath $a$}}
+ {\mbox{\boldmath $C$}})
 \cdot {\mbox{\boldmath $j$}} j^{-1}
+ M \right)}|{\mbox{\boldmath $e$}}_{i-1}>|{\mbox{\boldmath
$x$}}_{i-1}>
\nonumber \\
&=& \int \frac{d^3{\mbox{\boldmath $p$}}_i}{(2 \pi)^3}
<{\mbox{\boldmath $e$}}_i|{\mbox{\boldmath $e$}}_{i-1}>
\exp[-\Delta L \left( i({\mbox{\boldmath $p$}}_i
+ {\mbox{\boldmath $a$}}_i )
\cdot {\mbox{\boldmath $e$}}_i
+M \right)+ i{\mbox{\boldmath $p$}}
\cdot({\mbox{\boldmath $x$}}_i
-{\mbox{\boldmath $x$}}_{i-1})]
+ O(\Delta L^2)\;.\nonumber\\& &
\end{eqnarray}
Therefore, the free energy takes a form of path integrals
\begin{equation}
W_j=\int_{\epsilon}^{\infty} \frac{dL}{L}
\int {\cal D}{\mbox{\boldmath $x$}}
{\cal D}{\mbox{\boldmath $p$}}{\cal D}{\mbox{\boldmath $e$}}
\exp \left[ -\int_0^1 dt \left\{ i({\mbox{\boldmath $p$}}
+ {\mbox{\boldmath $a$}}
 + {\mbox{\boldmath $C$}}) \cdot {\mbox{\boldmath $e$}} L
+ M L - i{\mbox{\boldmath $p$}} \cdot \dot{{\mbox{\boldmath
$x$}}} \right\} +
ij \Xi [{\mbox{\boldmath $e$}}] \right],
\label{PP}
\end{equation}
where the parameter $t$ is rescaled as $t \rightarrow Lt$,
and $\dot{\mbox{\boldmath $x$}} = d{\mbox{\boldmath
$x$}}(t)/dt$.
Next, we try to integrate out ${\mbox{\boldmath $p$}}$,
${\mbox{\boldmath $e$}}$ and $L$. Let's map the
integral over $L$ to the integral over an einbein $h$,
which by definition is an one-form in one dimensional
space. Doing so, we take the advantage of the integral
thus being explicitly diffeomorphism invariant in the
one dimensional space.
In the Fadeev-Popov procedure, the integral over the
Schwinger parameter $L$ is regarded as a gauge fixed
path integral with respect to the
diffeomorphism transformation,
with $L$ the zero mode of the einbein. Namely, we have
the mapping
\begin{equation}
\lim_{\epsilon \rightarrow 0}
\int_\epsilon^{\infty} \frac{dL}{L}  \rightarrow
\int \frac{{\cal D}h}{V_{Diff}}\;.
\end{equation}
With this replacement, the path integral reads
\begin{equation}
W_j = \int \frac{{\cal D}h}{V_{Diff}} {\cal D}
{\mbox{\boldmath $x$}} {\cal D}{\mbox{\boldmath $p$}}
{\cal D}{\mbox{\boldmath $e$}} \exp \left[ -\int_0^1
\left\{ i( {\mbox{\boldmath $p$}} +
{\mbox{\boldmath $a$}} + {\mbox{\boldmath $C$}} )
\cdot {\mbox{\boldmath $e$}} h
 + M h - i{\mbox{\boldmath $p$}} \cdot
\dot{{\mbox{\boldmath $x$}}} \right\}
+ ij \Xi [{\mbox{\boldmath $e$}}] \right].
\end{equation}
Here, ${\mbox{\boldmath $p$}}$ is readily integrated out
\begin{equation}
W_j = \int \frac{{\cal D}h}{V_{Diff}}
{\cal D}{\mbox{\boldmath $e$}}
{\cal D}{\mbox{\boldmath $x$}}
\delta ( \dot{\mbox{\boldmath $x$}}
- {\mbox{\boldmath $e$}} h )
\exp \left[  - \int_0^1 dt \left\{
i ({\mbox{\boldmath $a$}} + {\mbox{\boldmath $C$}} )
\cdot {\mbox{\boldmath $e$}} h
+M h \right\} +i j \Xi [{\mbox{\boldmath $e$}}] \right].
\end{equation}
In this expression we see
$h = \sqrt{ {\dot{{\mbox{\boldmath $x$}}} }^2}$ and
\begin{equation}
{\mbox{\boldmath $e$}}(t)
= \dot{{\mbox{\boldmath $x$}}} /
\sqrt{ {\dot{{\mbox{\boldmath $x$}}}}^2} \label{e}\;,
\end{equation}
which is the unit tangent vector of the path parametrized
by the real variable $t$.
Carrying out the integrals over ${\mbox{\boldmath $e$}}$
and $h$,  finally we obtain
\begin{equation}
W_j = \int{\cal D}{\mbox{\boldmath $x$}}
\exp\left[ -\int^1_0dt \{ M\sqrt{\dot{{\mbox{\boldmath $x$}}}^2}
+i{\mbox{\boldmath $a$}}\cdot\dot{\mbox{\boldmath $x$}}
+i{\mbox{\boldmath $C$}}\cdot\dot{\mbox{\boldmath $x$}} \}
+i j \Xi[ \frac{\dot{{\mbox{\boldmath $x$}}}}
{|\dot{{\mbox{\boldmath $x$}}}|}]\right].
\label{WF}
\end{equation}
It is worth of notice that the free energies of
different spin-$j$ matter fields
differ from each other just by
 a coefficient, the value of spin $j$,
to the spin factor $\Xi$.

\vskip 0.5truein
\section{Spin Transmutation via Chern-Simons Term}
\setcounter{equation}{0}
\vspace{3 pt}

With $W_j$ in the form expressed in (\ref{WF}),
the integral over $a_\mu$ in the partition
function (\ref{Z1}) is readily to carry out,
it gives
\begin{equation}
Z(j,\alpha) =\sum^\infty_{n=1}\int\prod^n_{i=1}
  {\cal D}{\mbox{\boldmath $x$}}_i
\exp\left[-\sum^{n}_{i=1}\{
\int dt (M\sqrt{\dot{\mbox{\boldmath $x$}}_i^2}
+i{\mbox{\boldmath $C$}}\cdot\dot{\mbox{\boldmath $x$}}_i)
-i j \Xi[\frac{\dot{\mbox{\boldmath $x$}}_i}
{|\dot{\mbox{\boldmath $x$}}_i|}]
-i\frac{\alpha}{2}\Theta_{ii}\}
+i\alpha\sum_{i<k}\Theta_{ik}\right]\;,
\label{SFa}
\end{equation}
where ${\mbox{\boldmath $x$}}_i(t)$
is the position vector of the $i$th path, and
\begin{equation}
\Theta_{ik} =\frac{1}{\alpha}\int^1_0 dt\int^1_0 ds
\frac{dx_i^\mu}{dt}\frac{dx^\nu_k}{ds}
<a_\mu({\mbox{\boldmath $x$}}_i)
a_\nu({\mbox{\boldmath $x$}}_k)>\;.
\end{equation}
A convenient gauge choice is the Landau gauge, under which,
\begin{eqnarray}
 <a_\mu({\mbox{\boldmath $x$}})a_\nu({\mbox{\boldmath $y$}})>
&=&\int{\cal D}a_\lambda
a_\mu({\mbox{\boldmath $x$}}) a_\nu ({\mbox{\boldmath $y$}})
\exp[\frac{i}{8\pi\alpha}\int d{\mbox{\boldmath $x$}}
\epsilon^{\sigma\tau\eta}
a_\sigma\partial_\tau a_\eta]\nonumber\\
&=& 8\pi\alpha\epsilon_{\mu\nu\lambda}
\frac{x^\lambda-y^\lambda}{|{\mbox{\boldmath $x$}}-
{\mbox{\boldmath $y$}}|^3}\;.\label{CS}
\end{eqnarray}
Using notation of the unit vector
\begin{equation}
{\mbox{\boldmath $e$}}(s,t)
= \frac{{\mbox{\boldmath $x$}}_i(s)
-{\mbox{\boldmath $x$}}_k(t)}
{|{\mbox{\boldmath $x$}}_i(s)
-{\mbox{\boldmath $x$}}_k(t)|}\in S^2\;,
\label{ee}
\end{equation}
$\Theta_{ik}$ takes a compact form
\begin{equation}
\Theta_{ik}=\int^1_0ds\int^1_0 dt{\mbox{\boldmath $e$}}
\cdot[\partial_s{\mbox{\boldmath $e$}}
\times\partial_t{\mbox{\boldmath $e$}}]\;.
\label{theta}
\end{equation}
$\Theta_{ik}$ has been extensively discussed in ref. \cite{SSS}
and \cite{T},
here we present its main features with some comments.
First, if $i \neq k$, {\it i.e.}
the closed paths denoted by $i$ and $k$ do not coincide,
$\Theta_{ik}$, called relative energy,
is the Gauss linking number and thus is quantized \cite{IM}
\begin{equation}
\Theta_{ik} \in 4\pi{\mbox{\boldmath $Z$}}\;.
\label{psiij}
\end{equation}
 From (\ref{SFa}) and (\ref{psiij}), we see
that to any system that has an
integer or odd-half-integer
statistical parameter, $\alpha \in
{\mbox{\boldmath $Z$}}/2$,
the relative energy is irrelevant.
In other words, the contribution to the partition function
from the relative energy is associated with the `small'
portion of statistics $\alpha$ modular
integer and odd-half-integer.

Secondly, when the paths $i$ and $k$ do coincide,
$\Theta_{ii}$, called self-energy, is not quantized.
Instead it is related to the spin factor $\Xi$
(see (\ref{phi})) via
\begin{equation}
\Theta_{ii}-2\Xi = 4\pi ~~~(mod~~ 8\pi)\;.
\label{psiii}
\end{equation}
It seems when $i=j$ the definition (\ref{ee}) of
 ${\mbox{\boldmath $e$}}(t,s)$ 
 at the point $s=t$ is ambiguous.
To fix this problem, one could use the ``ribbon-splitting
technique'' \cite{T}, in which one splits the two coinciding
paths a little bit, and takes a wisely chosen limit
to merge them again later. Here we prefer a
resolution to well define the self-energy
by introducing a
limitation procedure without this splitting \cite{SSS}. Let's
change the variables of ${\mbox{\boldmath $e$}}(s,t)$ from
$(s,t)$ to $(u,t)$ with $u=s-t \in [0,1]$, and consider
the connection conditions at $u=0$ and $u=1$.
At $u=0$, one  naturally defines a limit
\begin{equation}
{\mbox{\boldmath $e$}}(u=0,t)= \lim_{\epsilon \rightarrow +0}
{\mbox{\boldmath $e$}}(\epsilon,t)
= {\mbox{\boldmath $e$}}(t) \equiv
\frac{\dot{\mbox{\boldmath $x$}}(t)}
{|\dot{\mbox{\boldmath $x$}}|}\;;
\end{equation}
then at $u=1$ it must be
 \begin{equation}
 {\mbox{\boldmath $e$}}(u=1,t) =\lim_{\epsilon \rightarrow +0}
 {\mbox{\boldmath $e$}}(1-\epsilon,t)
 =-{\mbox{\boldmath $e$}}(t)\;.
 \end{equation}
Therefore, ${\mbox{\boldmath $e$}}(u, t)$
satisfies the anti-periodic boundary
condition for the variable $u$
\begin{equation}
 {\mbox{\boldmath $e$}}(0,t)= -{\mbox{\boldmath $e$}}(1,t)
= {\mbox{\boldmath $e$}}(t)\;,
 \label{AB}
 \end{equation}
comparing the periodic boundary condition
for the variable $t$
 \begin{equation}
 {\mbox{\boldmath $e$}}(u,t+1)
={\mbox{\boldmath $e$}}(u,t)\;.
 \end{equation}
In this way, ${\mbox{\boldmath $e$}}$ and thus
the self-energy $\Theta_{ii}$ are well-defined on the whole
(smooth) path. 
Moreover, the factor $2$ on the left hand side
of (\ref{psiii}) reflects the fact that
the self-energy $\Theta_{ii}$ has two boundaries
at $u=0$ and $u=1$ while the spin factor $\Xi$
has only one. Due to the diffeomorphism invariance of
$\Theta_{ii}-2\Xi$ defined on a closed
path, one is free to deform the path continuously
onto a plane, where (\ref{psiii}) is readily to
obtain. Finally, we should mention that in this
approach, one assumes that,
when the random paths intersect,
a suitable point splitting can always be used
to regularize the intersecting point(s).

The quantization of the relative energy $\Theta_{ik}$, (\ref{psiij}),
and the simple relation of the self-energy $\Theta_{ii}$
with the spin factor, (\ref{psiii}), have very interesting
implication that the Chern-Simon coupling therefore can be
related to the spin of the matter field coupled.
This can be seen clearly from (\ref{SFa}).
Decreasing $\alpha$ by an integer or odd-half-integer
and increasing $j$ at the same time
by the same amount (or vise versa),
one doesn't change the partition
function at all. Namely, we have for any $n\in{\mbox{\boldmath $Z$}}$
\begin{equation}
Z(j,\alpha) = Z(j+ n/2,
\alpha -  n/2)\;. \label{AA}
\end{equation}
(\ref{AA}) is our main result here, which explicitly
exhibits the spin transmutation.
As a special case, when $\alpha$ is an integer
or odd-half-integer
(so that the relative energy is irrelevant),
one may transmute all the Chern-Simons coupling
into the spin of matter field by using (\ref{psiii}),
and obtain a free theory
of anyon with 
a spin  $j + \alpha$.

\vskip 0.5truein
\section{Equivalence of Chern-Simons Matter Models}
\setcounter{equation}{0}
\vspace{3 pt}

Treating anyons as fundamental fields
faces great difficulties
\cite{FF}.
More practical approach in dealing with anyons is to use
integer or odd-half-integer spin-$j$ fields
as the fundamental fields
and couple them to Chern-Simons fields. In this way,
traditional and powerful methods to solve interacting
field theories can be used directly.
In particular, to calculate quantum fluctuations,
perturbation expansion over the Chern-Simons coupling
$\sqrt{4\pi\alpha}$
is normally used. In many realistic systems,
unfortunately, the Chern-Simons interactions are
very strong, $i.e.$
$\alpha$ has to be so large \cite{note8}
that an expansion in
it is not well convergent.
With the spin transmutation mechanism seen here,
one can now trade a major part of Chern-Simons
coupling for higher spin. Then, one may have
an effective theory with a remaining small
Chern-Simons coupling.

For example, it is straightforward now to verify the
following three models being equivalent one to the
others, for any
given $\alpha$ \cite{CI}.
The first model has a spin-$1/2$ two-component Dirac
field $\psi$ as the fundamental field,
\begin{equation}
{\cal L}(\frac{1}{2}, \alpha) =  \bar\psi
[\gamma_\mu(\partial_\mu + ia_\mu +iC_\mu) +M]\psi
- \frac{i}{8\pi\alpha}
\epsilon_{\mu\nu\lambda}a_\mu\partial_\nu a_\lambda\;,
\label{SF}
\end{equation}
where the Dirac matrices in
three (Euclidean) dimensions $\gamma_{\mu}=\sigma_{\mu}$ with
$\sigma_{\mu}$ ($\mu=1,2,3$) Pauli matrices,
and $C_\mu$ again an external gauge field.
And the second involves
the spin-$1$ complex vector field $B_\mu$,
\begin{equation}
{\cal L}(1, \alpha-\frac{1}{2}) =
\frac{1}{2}B^*_{\mu}
[-i\epsilon^{\mu\nu\lambda}(\partial_\nu + ia_\nu+iC_\nu)
+M\delta^{\mu\lambda}]B_\lambda
- \frac{i}{8\pi (\alpha-1/2)}
\epsilon_{\mu\nu\lambda}a_\mu\partial_\nu a_\lambda\;.
\label{SB}
\end{equation}
Finally, the third contains a spin-$3/2$ four-component
fermion field $\Psi$
\begin{equation}
{\cal L}(\frac{3}{2},\alpha-1) =  \bar\Psi
[\frac{2}{3} L_\mu(\partial_\mu + ia_\mu +iC_\mu) +M]\Psi
- \frac{i}{8\pi (\alpha-1)}
\epsilon_{\mu\nu\lambda}a_\mu\partial_\nu a_\lambda\;,
\label{SFF}
\end{equation}
$L_\mu$ are $4\times4$ matrices satisfying the Lorentz algebra
and $L_\mu L_\mu = (3/2)(3/2+1)$.
Obviously
(\ref{SF}), (\ref{SB}), and (\ref{SFF}) can be good
perturbation theories only around
$\alpha = 0, 1/2,$ and $1$, respectively.
This claim of equivalence is based on (\ref{AA}) which takes
a form in the present case
\begin{equation}
Z(\frac{1}{2}, \alpha) = Z(1, \alpha - \frac{1}{2})
= Z(\frac{3}{2}, \alpha - 1) = \cdots \;.\label{BB}
\end{equation}
``$\cdots$'' here means  obvious extension of the equivalence
to ($\infty$ number) models with ``higher'' matter spins and
``weaker'' Chern-Simons interactions.
 Moreover, the equivalence can be generalized to some
matter field self-interactions. Let's take the four-body
interaction as an example.
Assume a local four-fermion interaction term
$(\bar{\psi}\psi)^2$ added to
the Chern-Simons Dirac fermion model (\ref{SF}),
the Lagrangian now reads
\begin{equation}
{\cal L}(\frac{1}{2},\alpha,g)
= \bar\psi [ \sigma_\mu (\partial_\mu + i a_\mu
+ i C_\mu ) + M] \psi +\frac{ g}{4} (\bar{\psi}\psi)^2
- \frac{i}{8 \pi \alpha} \epsilon_{
\mu \nu \lambda} a_\mu \partial_\nu a_\lambda\;,
\label{L3}
\end{equation}
where g is the coupling constant for the four-body interactions.

We introduce an auxiliary field $b(x) =i(g/2)\bar{\psi}\psi $ and turn
the four-body interaction into a Yukawa interaction:
\begin{equation}
{\cal L}_b(\frac{1}{2},\alpha,g)
= \bar\psi [ \sigma_\mu (\partial_\mu + i a_\mu
+ i C_\mu ) + M + ib ] \psi - \frac{i}{8 \pi \alpha} \epsilon_{
\mu \nu \lambda} a_\mu \partial_\nu a_\lambda
+ \frac{1}{g}b^2\;.
\end{equation}
The four-body interaction in (\ref{L3}) is
recovered by performing the Gaussian functional
integral over $b$ in the partition function
\begin{equation}
Z(\frac{1}{2},\alpha,g) = \int{\cal D}b{\cal D}a_\mu{\cal D}
\bar{\psi}
{\cal D}\psi e^{-\int{\cal L}_b(1/2,\alpha,g)}\;.
\end{equation}
Integrating out the Dirac fields, we have the free energy
for spin-$1/2$ field
\begin{equation}
W_{1/2} = \int{\cal D}{\mbox{\boldmath $x$}}
\exp\left[-\int^1_0 dt\{ \left(M + ib({\mbox{\boldmath $x$}})
\right)
\sqrt{\dot{{\mbox{\boldmath $x$}}}^2}
+i{\mbox{\boldmath $a$}}\cdot\dot{\mbox{\boldmath $x$}}
+i{\mbox{\boldmath $C$}}\cdot\dot{\mbox{\boldmath $x$}}\}
+ \frac{i}{2} \Xi[ \frac{\dot{{\mbox{\boldmath $x$}}}}
{|\dot{{\mbox{\boldmath $x$}}}|}]\right]\;.
\label{Wb}
\end{equation}
Compared to  (\ref{WF}) (taken
$j=1/2$), which involves no self-interaction,
in (\ref{Wb}), the change is only to
replace the mass term $M$ with
$M+b({\mbox{\boldmath $x$}})$.
Then, carrying out
the integral over $a_\mu$ as done in the previous section,
we obtain
\begin{eqnarray}
Z(\frac{1}{2},\alpha,g) &=& \sum^\infty_{n=1} \frac{1}{n!}
\int {\cal D}b
\exp[ -\int d^3x b^2/g ]
\int \prod^n_{i=1}
  {\cal D}{\mbox{\boldmath $x$}}_i \nonumber \\
& &\exp\left[-\sum^{n}_{i=1}[\int dt \{ (M
+ b({\mbox{\boldmath $x$}}))
\sqrt{\dot{\mbox{\boldmath $x$}}_i^2}
+ i{\mbox{\boldmath $C$}}\cdot\dot{\mbox{\boldmath $x$}}_i\}
- \frac{i}{2} \Xi[\frac{\dot{\mbox{\boldmath $x$}}_i}
{|\dot{\mbox{\boldmath $x$}}_i|}]
-i\frac{\alpha}{2}\Theta_{ii}]
+i\alpha\sum_{i<j}\Theta_{ik}\right]\;.\nonumber\\
& &
\label{SFb}
\end{eqnarray}
Now, we use (\ref{psiij}) and (\ref{psiii}) to shift $\alpha$
by an integer or odd-half-integer. For any $n\in{\mbox{\boldmath $Z$}}$,
we obtain
\begin{equation}
Z(\frac{1}{2},\alpha,g) =
Z(\frac{1+n}{2},\alpha
- \frac{n}{2},g)\;.
\end{equation}
Taking $n=1$ as an example, the right hand side above,
$Z(1,\alpha -\frac{1}{2},g)$, is the partition function
of the vector Chern-Simons model with four-body interaction
\begin{equation}
{\cal L}(1,\alpha-\frac{1}{2},g)
= - i \epsilon_{\mu \nu \lambda} B^*_\mu ( \partial_\nu
+ i a_\nu + i C_\nu +M ) B_\lambda + \frac{g}{4}(B^*_\mu B_\mu)^2
- \frac{i}{8 \pi (\alpha - 1/2)} \epsilon_{\mu \nu \lambda}
a_\mu \partial_\nu a_\lambda\;. \label{L4}
\end{equation}
Now, we have seen the equivalence between the models
(\ref{L3}) and (\ref{L4}).

It is straightforward to play this game over
again for more versions of the Chern-Simons
matter model for any given $\alpha$.

\vskip 0.5truein
\section{Concluding Remarks}
\setcounter{equation}{0}
\vspace{3 pt}

In this paper, we have shown explicitly how the
spin of particles transmutes when they interact
to the Chern-Simons gauge field. Doing so, we
have uncovered the intrinsic relation, the equivalence,
of a large class Chern-Simons matter models.
An immediate consequence of the equivalence
is that one is free to
select one of the equivalent anyon models
(though they may look like very different) that is
most suitable for exploring the features of the anyon
system one is interested in.

As the higher-spin matter fileds in general
carry more degrees of freedom,
there seems a problem with a counting of degrees of freedom in
equivalent models. The answer is that the Chern-Simons interaction
affects the number of degrees of freedom of the whole anyon system.
As is seen in section 4, an integer or odd-half-integer
part of the Chern-Simons coupling $\alpha$ can be reabsorbed
by changing the spin of matter field.
This change of spin is accompanied by a change of numbers of
degrees of freedom carried by the matter field.
What happens here is a transfer but not creation of degrees
of freedom. Since equivalent models can be derived
from each other, all these have the same
amount degrees of freedom and
describe the same many-body system with arbitrary $\alpha$.

Besides the statistics and spin, one more interesting
character of anyons is the scaling dimension, which governs
the asymptotical behavior of anyon systems.
Due to the Fermi-Bose transmutation, one is led
to expect the scaling dimensions of anyons be dependent of
the statistics parameter, $i.e.$ $d = d(\alpha)$.
Take the Chern-Simons Dirac field model (\ref{SF}) as an example.
When $\alpha$ is an integer or odd-half-integer the model describes
free fermions or bosons, respectively.
There must be for instance $d = d(\alpha = 0) = 1$ at the Dirac
fermion point and
$d=d(\alpha=-1/2) = 1/2$ at the scalar point, since for free
fields the scaling dimensions are just the
engineering dimensions.
However, when $\alpha$ varies away from integer and odd-half-integer,
the scaling dimension of anyon should be a sum of
the engineering dimension of fundamental field and
the $\alpha$ dependent correction from
Chern-Simons interaction. This is again a quantum-level phenomenon --
the asymptotical behavior of a system is usually changed by
quantum fluctuations.
In renormalizable field theories,
a correction to the scaling
dimension of field, called an anomalous dimension, is
computed normally by the renormalization group method.

The calculations to two-loop \cite{CSW}
show that the anomalous dimensions
of the matter fields are a continuous decreasing function
of $\alpha^2$ \cite{note5}.
The perturbation results are reliable near
$\alpha = 0$,
but less and less reliable as $|\alpha|$
goes larger and larger.
As an example in applying the equivalence,
the authors of \cite{CI}
calculated the anomalous dimension of
the matter field in a
system where each fermion particle
carries about one flux tube,
$i.e.$ the model (\ref{SF}) with
$\alpha \sim 1/2$, which reflects
a strong coupling. The calculation then
is done by mapping (\ref{SF}) to the
Chern-Simons field coupled to the vector matter model (\ref{SB}).
The anomalous dimension of the vector
matter field is calculated as
a decreasing function of $(\alpha-1/2)^2$,
which is well convergent near $\alpha=1/2$ \cite{note10}.

There has been debate on whether the only effect
of Chern-Simons gauge field is to endow the
particle with arbitrary spin
or whether residual interactions are present \cite{HKP}. We are now
able to shed some lights on this issue. Looking at the partition
function (\ref{SFa}) and the conditions (\ref{psiij}) and
(\ref{psiii}), it is clear that only the integer
or odd-half-integer part of the Chern-Simons coefficient can be absorbed
into the spin of the fundamental field. In other words, our analysis
here suggests that:
only when $\alpha$ is an integer or odd-half-integer, the sole role
the Chern-Simons field plays is to change the spin of particle by
an amount $\alpha$; otherwise, a residual Chern-Simons interaction,
minimally the part of $\alpha$ that is less than $1/2$, must be
present.
The complication is apparently due to the relative energy
$\Theta_{ik}$, which is
quantized in a way shown in (\ref{psiij}).

We conclude this paper with remarks on two more issues.
The relativistic anyon theory we have considered,
(\ref{SJ}), is rather general as it represents
infinite number models characterized by two parameters
$j$ and $\alpha$.
However, this model doesn't admit the scalar
as a fundamental field, as $j \neq 0$.
A local linear scalar theory is somehow special
in the sense that its kinetic term must involve a second derivative.
On the other hand, based on the equivalence discussed
here, it is possible to construct a kind of
spin-$0$ field theory by coupling spinning fields
to Chern-Simons field with
$j+\alpha = 0$ for exactly free boson particles,
or $=\lambda \sim 0$ for boson particles near free.
A sample model is that the (spin-$1/2$) Dirac field couples
to Chern-Simons field with the coupling $\alpha \sim -1/2$.
It is worth of notice that, since the fundamental Dirac
field obeys the Pauli exclusion principle,
so do the constructed spin-$0$
fields. The particles described by such a theory are
known as hard core bosons.

One might be tempted to understand
whether the class of
models we discussed here can serve as a critical model
that describes the critical behavior or phase transition
of anyon systems. For $j=1/2$, {\it i.e.} for the
Chern-Simons
Dirac fermion model, it is indeed the case. In
ref. \cite{CFW}, the authors analyzed a lattice model of
anyons in a periodic potential and an external magnetic
field which exhibits a second order transition from a
Mott insulator to a quantum Hall fluid.
The continuum limit of this lattice model is shown exactly
the Chern-Simons Dirac fermion model (\ref{SF})
(with $C_\mu$ missed), and the transition
is characterized by the anyon statistics, $\alpha$. We
would guess that all odd-half-integer $j$ models are in this nature.

On the other hand, however, for $j=1$, {\it i.e.} for the
Chern-Simons vector boson model (\ref{SB}),
 and likely for
all the integer $j$ models, there appears a kind of discontinuity
when the mass parameter $M$ is taken to be zero.
In the model (\ref{SB}),
the only degree of freedom is carried by the mass term of
the vector field (so this degree of
freedom is of longitudinal).  When the mass parameter
$M = 0$ is taken, the only degree of freedom
is missing and the
U(1) Chern-Simons (vector) matter theory becomes
a topological field theory. To see this,
we set
\begin{equation}
 a_\mu=eA^1_\mu,~~ {\rm and} ~~B_\mu = A_\mu^2+iA_\mu^3\;,
\end{equation}
where $e=\sqrt{4\pi(\alpha-1/2)}$, and substitute these
into (\ref{SB}) with $C_\mu = 0$. Then we obtain
\begin{equation}
{\cal L} = -\frac{i}{2}\epsilon_{\mu\nu\lambda}
(A^a_\mu\partial_\nu A^a_\lambda
+\frac{e}{3}\epsilon^{abc}A^a_\mu A^b_\nu A^c_\nu)\;,
\label{to}
\end{equation}
upto a total derivative term. (\ref{to}) is the well-known $SU(2)$
Chern-Simons field theory, it is independent of a choice of the
space-time metric.

The seeming singular behavior of the spin-1 model at $M=0$ raises
a question whether the spin-1 field theory exists as a
continuum limit of a lattice model? This question can be answered
only by studying the corresponding lattice model.
A research
of the quantum diffusion process with the spin
factor as a free field theory on three
dimensional lattice
suggests there exists a continuum
limit for
the spin-1 model, as it does for the spin-1/2 model
\cite{CH}. Meanwhile, the research of
diffusion process
also shows that $M=0$ is
indeed a singular point for spin-1 model.
This explains why,
as continuum field
theories, the two models describe phase transitions
of different
nature. As $M \rightarrow 0$, in the spin-1 model
it restores
topological and non-Abelian gauge invariance
\cite{chen},
but in the spin-1/2 model
it concerns only a global conformal symmetry.
Obviously, the equivalence between the spin-1
and spin-1/2 models can not be
simply extrapolated to the $M=0$
case.

The authors thank A. Kovner, G. Semenoff, H. Van Dam,
and Y.-S. Wu for discussions.
The work of W.C. was supported in part by
the  U.S. DOE under the contract No. DE-FG05-85ER-40219.

Notice: after this work was finished, we
noticed ref. \cite{Fo} where a similar issue
was discussed in a different way.


\baselineskip=18.0truept
\begin{thebibliography}{199}
\bibitem{M}
E. Merzbacher, Am. J. Phys. {\bf 30}, 237 (1962);\\
J. Leinaas and J. Myrheim, Nuovo Cimento {\bf 37}, 1 (1977);\\
F. Wilczek, Phys. Rev. Lett. {\bf 49}, 957 (1982);\\
For a modern review, see F. Wilczek, {\it Fractional Statistics
and Anyon Superconductivity} (World Scientific, Singapore, 1990).
\bibitem{H}
C. Hagen, Ann. Phys. (N.Y.) {\bf 157}, 342 (1984);
Phys. Rev. D {\bf 31}, 848 (1985); {\bf 31}, 2135 (1985);\\
D. Arovas, J. Schrieffer, F. Wilczek, and A. Zee, Nucl. Phys.
{\bf B251}, 117 (1985).
\bibitem{P}
A.M. Polyakov, Modn. Phys. Lett. {\bf A3}, 325 (1988);\\
C.-H. Tze, Int. J. Mod. Phys. {\bf A3} 1959 (1988);\\
S. Iso, C. Itoi, and H. Mukaida, Nucl. Phys. {\bf B346}, 293 (1990);\\
N. Shaju, R. Shankar, and M. Sivakumar, Modn. Phys. Lett. {\bf A5},
593 (1990).
\bibitem{KS}
I. Kogan, Phys. Lett. {\bf 262B}, 83 (1991);\\
I. Kogan and G. Semenoff, Nucl. Phys. {\bf B368}, 718 (1991).
\bibitem{CI}
W. Chen and C. Itoi,
Phys. Rev. Letts, {\bf 72}, 2527 (1994).
\bibitem{Sch}
It was Schonfeld who first observed that the geometry of
a $(2+1)$ dimensional space-time allows for fractional
eigenvalues of spin in Nucl. Phys. {\bf B185}, 157 (1981);
see also \cite{JN} \cite{KS}.
\bibitem{Pl} M.S. Plyushchay, Phys. Lett. {\bf B262}, 71 (1991); Nucl. Phys.
{\bf B362}, 54 (1991).
\bibitem{B}
B. Binegar, J. Math. Phys. {\bf 23}(8), 1511 (1982).
\bibitem{note6}
This mass spectrum
reflects in the particle propagator as
multi-poles at $p^2=(jM/k)^2$.
\bibitem{JN}
R. Jackiw and V.P. Nair,
Phys. Rev. D {\bf 43}, 1933 (1991).
\bibitem{note1} The external gauge field $C_\mu$, if it is taken as
a magnetic monopole, affects the spin of
matter fields too.
We shall not discuss this further in the present paper.
\bibitem{note3} More precisely, the sign of mass differs the
right- from left-hand moving spin, as the particles in $(2+1)$
dimensions are actually circularly polarized, and there are no symmetric
transformations to relate the spin-``up'' and ``down'' states,
unlike in the $(3+1)$ dimensions.
\bibitem{SSS}
see S. Iso {\it et. al.} and N. Shaju {\it et. al.} in \cite{P};\\
also see T. Jaroszewicz and P.S. Kurzepa,
Ann. Phys. (N.Y.) {\bf 210}, 255 (1991);
 {\bf 213}, 135 (1992);
 {\bf 216}, 226 (1992).
\bibitem{AP}
J.R. Klauder and B.S. Skagerstam,
{\it Coherent States, Applications in
Physics and Mathematical Physics}
(World Scientific, Singapore, 1985);\\
A. Perelomov,
{\it Generalized Coherent States and Their Applications}
(Springer-Verlag, Berlin, 1986);\\
E. Fradkin and M. Stone, Phys. Rev. B {\bf 38},
7215 (1988);\\
E. Fradkin, {\it Field Theories of Condensed Matter Systems},
(Addison \& Wesley, 1991).
\bibitem{T}
E. Witten, Commun. Math. Phys. {\bf 92}, 451 (1984);
{\bf 121}, 351 (1989);\\
C.H. Tze, in \cite{P}.
\bibitem{IM}
see also
I. Kogan and A. Yu. Morozov, Sov. Phys. JETP {\bf 61} 1 (1985).
\bibitem{FF}
For instance, see F. Wilczek
(1990) in \cite{M};\\
M.S. Plyushchay, Phys. Lett. {\bf B248}, 107 (1990); {\bf B273}, 250 (1991);
{\bf B320}, 91 (1993); Inter. J. Modn. Phys. {\bf A7}, 7054 (1992).
\bibitem{note8}
For example, in a recent successful theory of the half-filled
Landau level \cite{HLR}, each electron carries about two flux
tubes, that corresponds to $\alpha \sim 1$ in the
Chern-Simons Dirac fermion model.
\bibitem{HLR}
B.I. Halperin, P.A. Lee, and N. Read, Phys. Rev. B {\bf 47}
 7312 (1993);\\
V. Kalmeyer and S.-C. Zhang, {\it ibid.} {\bf 46}
 9889 (1992).
\bibitem{CSW}
W. Chen, G.W. Semenoff, and Y.-S. Wu, Phys.
Rev. D {\bf 44}, R1625 (1991);
{\it ibid} {\bf 46}, 5521 (1992);
W. Chen and M. Li, Phys. Rev. Lett. {\bf 70}, 884 (1993).
\bibitem{note5} The $\beta$ function of the
Chern-Simons coupling $\alpha$
is identically vanishing. Therefore,
$\alpha$ serves well as a controlling
parameter in a perturbative expansion.
\bibitem{note10} In this calculation, we
took a higher energy limit ($M/|p| \rightarrow 0$)
for simplicity.
\bibitem{HKP} J. Hong, Y. Kim and P.Y. Pac, Phys. Rev. Lett. {\bf 64},
2230 (1990);\\
R. Jackiw and E. Weinberg, {\it ibid}. {\bf 64}, 2234 (1990);\\
R. Jackiw and S.-Y. Pi, {\it ibid.} {\bf 64}, 2969 (1990);
Phys. Rev. D{\bf 42}, 3500 (1990);  \\
R. Jackiw, K. Lee and E. Weinberg, {\it ibid.} {\bf 42}, 3488 (1990).
\bibitem{CFW}
W. Chen, M.A.P. Fisher, and  Y.-S. Wu, Phys. Rev. B
{\bf 48}  13749 (1993).
\bibitem{CH}
C. Itoi, (under preparation).
\bibitem{chen}
 W. Chen, {\it Phase Transition in $(2+1)d$ Quantum Gravity},
IFT-490-UNC (1994).
\bibitem{Fo}
S. Forte, Int. J. Modn. Phys. {\bf A7}, 1025 (1992);\\
S.Forte, T. Jolicoeur, Nucl. Phys. {\bf B350}, 589 (1991).
\end{thebibliography}
\end{document}